\documentclass[]{spie}  %>>> use for US letter paper
\usepackage[]{graphicx}
\usepackage{amsmath}

\pdfoutput=1
\usepackage[pdftex, bookmarks=true, bookmarksnumbered=true, plainpages=false, breaklinks=true, colorlinks=true, urlcolor=blue, citecolor=blue, linkcolor=blue]{hyperref}

\newcommand{\be}{\begin{equation}}
\newcommand{\ee}{\end{equation}}
\newcommand{\beN}{\begin{equation*}}
\newcommand{\eeN}{\end{equation*}}

\newcommand{\Dlambda}{{\Delta\lambda}}

\newcommand{\SR}{\mathcal{R}} %Spectral Resolution
\newcommand{\St}{\mathcal{S}} %Strehl ratio

\title{Astrophotonic micro-spectrographs in the era of ELTs}

\author{N. Blind \supit{a, b}, E. Le Coarer \supit{b}, P. Kern \supit{b}, J. Bland-Hawthorn \supit{c,d}
\skiplinehalf
\supit{a} Max-Planck-Institut f\"ur extraterrestrische Physik, Giessenbachstr.1, D-85748 Garching, Germany; \\
\supit{b} Institut de Plan\'etologie et Astrophysique de Grenoble (Universit\'ee J. Fourier, CNRS, UMR 5571), BP 53, 38041 Grenoble, France;\\
\supit{c} Sydney Institute for Astronomy (SIfA), School of Physics, University of Sydney, NSW 2006, Australia;\\
\supit{d} Institute of Photonics and Optical Science (IPOS), School of Physics, University of Sydney, Australia\\
}

\authorinfo{blind@mpe.mpg.de}
\begin{document}

\maketitle

\begin{abstract}
The next generation of Extremely Large Telescopes (ELT), with diameters up to 39 meters, will start operation in the next decade and promises new challenges in the development of instruments. The growing field of astrophotonics (the use of photonic technologies in astronomy) can partly solve this problem by allowing mass production of fully integrated and robust instruments combining various optical functions, with the potential to reduce the size, complexity and cost of instruments. In this paper, we focus on developments in integrated micro-spectrographs and their potential for ELTs. We take an inventory of the identified technologies currently in development, and compare the performance of the different concepts. We show that in the current context of single-mode instruments, integrated spectrographs making use of, e.g., a photonic lantern can be a solution to reach the desired performance. However, in the longer term, there is a clear need to develop multimode devices to improve overall the throughput and sensitivity, while decreasing the instrument complexity.
\end{abstract}

\keywords{astrophotonics, photonics, spectrographs, ELT}

\section{Introduction} % (fold)
\label{part:introduction}

Spectroscopy plays a major role in astronomy with applications ranging from the study of celestial dynamics to cosmology. The next generation of Extremely Large Telescopes is planned to begin operation in the next decade and promises significant breakthroughs in our understanding of the Universe, as well as new challenges on the instrument design: the size of the instruments increases in proportion of the telescope diameter $D$, while their cost increases at least as $D^2$ \cite{bland_2006a}. This will lead to instruments whose cost are comparable to current 8\:m class telescopes \cite{russel_2004a, allington_2010b}

Astrophotonics is a new research field that lies at the interface of astronomy and photonics \cite{bland_2012a}. Fibers were used since the 70s to transport light from the telescope focus to a spectrograph, but astrophotonics also provides lots of original optical functions with applications in:
\begin{itemize}
\item Spectroscopy with fiber grating OH-filters \cite{bland_2009a} and integrated spectrographs (focus of this proceeding);
\item High-angular resolution with beam combiners for long baseline interferometry \cite{benisty_2009} and pupil remapping \cite{huby_2012a, jovanovic_2012a};
\item Mode conversion in waveguides with the photonic lanterns \cite{leon-saval_2005a, noordegraaf_2009a}.
\end{itemize}
Most of these technologies have already been demonstrated on-sky. In this proceeding, we focus on astrophotonic spectrographs with a view to achieving certain goals:
\begin{itemize}
\item Conceive fully integrated instruments, with the promise for more compact, more simple, and more stable systems than classical bulk optics;
\item Minimize useless pixels thanks to a better/easier arrangement of light into the detector (conversely to, e.g., lenslet and masked multi-object spectrographs like TIGER \cite{bacon_1988a});
\item Allow mass production of optical components at a lower cost for the future highly multiplex instrumentation.
\end{itemize}
Astrophotonics has the potential to provide the solution to the continuous increase in complexity of future optical/infra-red ground-based spectrographs, while reducing both size and cost of the instruments. The compactness of these instruments is one of the most obvious advantages in the context of space missions, and these have already found important applications on rockets, balloons and drones \cite{fogarty_2012a}.

%The paper is organized as follow. In Sect.~\ref{part:group_spectro}, we recall some basics of spectroscopy, while Sect.~\ref{part:optical_etendue} focuses on the notion of optical \'etendue. Then we focus on the integrated spectrometers in Sect.~\ref{part:identified_technos}, by defining first a series of characteristics and merit criteria, allowing to compare the identified technologies presented next. Performances of the different families of integrated spectrographs are finally compared in Sect.~\ref{part:signal_to_noise}, in the case of typical astrophysical sources and observing conditions. In Sect.~\ref{part:conclusion}, we summarise our reasons for believing that integrated spectrographs will be more widely used in the near future.

%%%%%%%%%%%%%%%%%%%%%%%%%%%%%%%%%%%%%%%%%%%%%%%%%%%%%%%%%%%%
%%%%%%%%%%%%%%%%%%%%%%%%%%%%%%%%%%%%%%%%%%%%%%%%%%%%%%%%%%%%
%%%%%%%%%%%%%%%%%%%%%%%%%%%%%%%%%%%%%%%%%%%%%%%%%%%%%%%%%%%%

%%%%%%%%%%%%%%%%%%%%%%%%%%%%%%%%%%%%%%%%%%%%%%%%%%%%%%%%
\begin{table*}[b]
\caption[]{Number of modes $M \sim S\Omega/\lambda^2$ of point sources according to the expected performance of MCAO/LTAO systems on the 42-m E-ELT (former design) in standard conditions (seeing = 0.8'') \cite{diolati_2010b, fusco_2010c}. EE stands for Encircled Energy. \smallskip \label{tab:strehl_ELT}}
\begin{center}
\begin{tabular}{llrrrrrrr}
\hline \hline
Band && V & I & J & H & K & M & N\\
\hline
%\multicolumn{9}{l}{Optical \'etendue S$\Omega/\lambda^2$}\\
Point + AO ($EE > 50$\%) && $> 1500$    & $> 500$ & 250        & 12          &  3            & 2         & 1 \\
Point + AO ($EE > 85$\%) && $\gg 1500$ & $> 500$ & $>250$ & $>150$  &  $> 80$  & $>16$ & $> 4$ \\
Resolved ($\theta =50$mas) && 375 & 125 & 62 & 38 & 20 & 4 & 1 \\
Resolved ($\theta =100$mas) && 1500 & 500 & 250 & 150 & 78 & 16 & 4 \\
\hline
\end{tabular}
\end{center}
\end{table*}
%%%%%%%%%%%%%%%%%%%%%%%%%%%%%%%%%%%%%%%%%%%%%%%%%%%%%%%%

\section{Photonic spectrometers} % (fold)

The Wiener-Khinchin theorem states that the power spectral density of a stationary random process is the Fourier transform of the corresponding autocorrelation function. It is the basis of spectrometry, and can be summarized as an autocorrelation of light propagation in time. This autocorrelation can be obtained with at least three different concepts: 
\begin{itemize}
\item Diffracting elements, like gratings, which are the most common type of spectrographs in operation today;
\item Interferential filters like Fabry-P\'erots, where only few specific spectral regions are transmitted;
\item Fourier Transform Spectrographs, where we measure the signal autocorrelation, and the Fourier transform operation is performed numerically.
\end{itemize}
A fourth way is now under investigation with energy sensitive detectors. They are not considered in the following.

Today, astrophotonics technologies are mostly single-mode (SM) devices, i.e we can couple at best a source with an optical etendue $S\Omega_{SM} = \lambda^2$. This corresponds roughly to the diffraction limit of a telescope (whatever its size): in this case, the coupling efficiency reaches a maximum value of 82\%. But in most cases, the source is spatially extended, either because it is spatially resolved (e.g.~galaxies or nebulae), and/or because of the atmospheric turbulence that spreads the stellar light into a large speckle pattern. Although in the latter case, Adaptive Optics systems can significantly reduce the etendue for point sources, $S\Omega$ is still high as long as we do not make use of Extreme AO systems. In all other situations, the optical etendue can reach values up to several $1000 \:\lambda^2$ for ELTs, especially in visible wavelengths (Tab.~\ref{tab:strehl_ELT}), i.e.~only $\sim$1/1000 of the light gathered by the telescope is actually coupled to the instrument.

Thanks to the high number of modes $M$ that Multi-mode (MM) waveguides can accept, the optical etendue of the system is greatly improved ($S\Omega \sim M \lambda^2$). However, considering a final photonic spectrograph --mostly SM devices in the current framework--, the coupling from the MM waveguide to the spectrograph will be similar to the direct injection into a SM waveguide, i.e. very low. MM waveguides can in addition limit spectral resolution because of modes propagating at different speeds, then generating a series of shifted and stretched spectra on the detector.

The photonic lantern\cite{leon-saval_2005a, noordegraaf_2009a} becomes a key component in this context: by converting an initial MM beam into multiple SM ones, the initial etendue can be preserved and we make a correct use of the large aperture of the telescope. Then each SM output can feed efficiently a different spectrograph. We refer to such devices as Multi Single-Mode (M-SM) hereafter. Such spectrographs actually offer an efficient way to overcome the Jacquinot criterion: each waveguide collecting a beam with an elementary optical \'etendue, the spectral resolution is not limited by the source size anymore. This is a unique property of these devices, allowing very high spectral resolution spectroscopy on very extended sources\cite{betters_2013a}.

%%%%%%%%%%%%%%%%%%%%%%%%%%%%%%%%%%%
\begin{table*}[b]
	\caption{Summary of identified grating/disperser technologies. \label{tab:gratings_comparison}}
	\begin{center}
	\begin{tabular}{llllll}
	\hline \hline
	 & \bf{CGS}\cite{grabarnik_2007a} & \bf{AWG-Vis}\cite{bland_2006a} & \bf{AWG-NIR}\cite{bland_2006a} & \bf{SHD}\cite{avrutsky_2006a} & \bf{PCS}\cite{lupu_2004a, momeni_2009a} \\
	\hline
	Spectral domain [nm]& 400-800 & 780-920 & 1465-1785 & 400-700 & 1560-1610\\
	Spectral resolution $\SR$ & 150-200 & 150-250 & 2000 &800 & 500\\
	Modicity & SM, MM & SM, M-SM & SM, M-SM & SM, M-SM & SM\\
	%$S\omega$  & 400 & 1 & 1 & 1 & 1 \\
	Simultaneous spectra & 1 & 12 & 12& 35 & 1\\
	%Reduced volume $m$ & $5\cdot 10^{6}$& $3\cdot 10^{9}$& $6\cdot 10^{8}$ & $8\cdot 10^{9}$ & $3 \cdot 10^{5}$ \\
	%%Operability & $\frownie$ & $\smiley$ & $\smiley$ & $\frownie$ & $\smiley$ \\
	%Efficiency & 0.3 & 0.6 & 0.5 & 0.7 & 0.1 \\
	%Sampling & parallel & parallel & parallel & parallel & parallel\\
	%Polarization & 1 pol & 1 pol & 1 pol & 1 pol  & 1 pol \\
	3D multiplexing & Bad & Possible & Possible & Bad & Possible\\
	%TRL & 4 & 2 & 6  & 4 & 3\\
	\hline
\end{tabular}
\end{center}
\end{table*}
%%%%%%%%%%%%%%%%%%%%%%%%%%%%%%%%%%%

%%%%%%%%%%%%%%%%%%%%%%%%%%%%%%%%%%%%%%%%%%%%%%%%%%%%%%%%%
\begin{table*}[t]
	\caption{Summary of identified Fabry-P\'erot and filter technologies. \label{tab:FP_comparison}}
	\begin{center}
	\begin{tabular}{lllll}
	\hline \hline
	 & \bf{ZZS}\cite{murry_2001a} & \bf{MEMS}\cite{wolffenbuttel_2005a}& \bf{PPSI}\cite{laux_2008a}  \\
	\hline
		Spectral domain [nm] & 1300-1800 & 1000-1785 & 400-700 \\
		Spectral resolution $\SR$ & 50 & 100 &5 \\
		Modicity & SM, MM & MM & MM \\
		%$S\omega$ & 45 & 1 &100& 10 \\
		Simultaneous spectra&  1 & 1000& $10^{6}$ \\
		%Reduced volume $m$ & $6\cdot 10^{6}$& $7\cdot 10^{7}$& $3\cdot10^3$ & $1\cdot 10^4$ \\
		%%Operability & $\smiley$ & $\smiley$ & $\smiley$ & $\smiley$  \\
		%Efficiency & 0.7 & 0.3 & 0.5 & 0.1 \\
		%Sampling & parallel & parallel & parallel & parallel \\
		%Polarization & 2 pol & 2 pol & 2 pol & 2 pol   \\
		3D multiplexing & Good & Good & Very good \\
		%TRL & 9 & 2 & 3/4 & 3 \\
%		Astrophysical targets & No ! & Compact & Wide field & UBV Photometry\\
		\hline
\end{tabular}
\end{center}
\end{table*}
%%%%%%%%%%%%%%%%%%%%%%%%%%%%%%%%%%%%%%%%%%%%%%%%%%%

%%%%%%%%%%%%%%%%%%%%%%%%%%%%%%%%%%%%%%%%%%%%%%%%%%%
\begin{table*}[h!]
	\caption[]{Summary of identified FTS technologies. 	\label{tab:FTS-compare}}
	\begin{center}
	\begin{tabular}{lllllll}
	\hline \hline
		 & \bf{$\mu$SPOC}\cite{rommeluere_2007a} & \bf{SWIFTS}\cite{lecoarer_2007a}& \bf{LLIFTS}\cite{martinB_2009b}  & \bf{AMZI}\cite{florjanczyk_2007a}\\
		\hline
		Spectral domain [nm]& 3300-5000 & 400-1000 & 1200-1900 & \\
		Spectral resolution $\SR$ & 1000 & 100000 &256 & 3600 \\
		Modicity & MM & SM, M-SM & SM & M-SM \\
		%$S\omega$ & 256 & 1 & 1-256 & 1 & 50-200 \\
		Simultaneous spectra& $>$ 300  & 1& 1 & 1 \\
		%Reduced volume $m$ & $2\cdot 10^{3}$& $5\cdot 10^{5}$& $3\cdot 10^{5}$ & $3\cdot 10^{7}$ &$3\cdot 10^8$\\
		%%Operability & $\smiley$ & $\smiley$ & $\smiley$  & $\smiley$ & $\frownie$\\
		%Efficiency & 0.2 & 0.1 & 0.3 & 0.7 & 0.5 \\
		%Sampling & Static & Static & Static & Static & Static \\
		%Polarization & 2 pol & 1 pol & 1 pol & 1 pol &  1 pol \\
		3D multiplexing & No & Possible & Bad & Possible \\
		%TRL & 5 & 4 & 9 & 4 & 4 \\
%		Astrophysical targets & Compact & Compact & Compact & Compact & Compact\\
%                                  & Faint, wide &            &                  &               & Faint\\
		\hline
\end{tabular}
\end{center}
\end{table*}
%%%%%%%%%%%%%%%%%%%%%%%%%%%%%%%%%%%%%%%%%%%%%%%%%%%%
%

%%%%%%%%%%%%%%%%%%%%%%%%%%%%%%%%%%%%%%%%%%%%%%%%%%%%%%%%%%%%
%%%%%%%%%%%%%%%%%%%%%%%%%%%%%%%%%%%%%%%%%%%%%%%%%%%%%%%%%%%%
%%%%%%%%%%%%%%%%%%%%%%%%%%%%%%%%%%%%%%%%%%%%%%%%%%%%%%%%%%%%

\section{\bf Identified technologies}
\label{part:merit_criteria of integrated spectrometers}

We can define a series of merit criteria based on practical considerations that help to sort and compare the different concepts and technologies, among which:
\begin{itemize}
\item {\bf Multiplexing capability} -- With astrophotonics, there are two different ways to access to multiplexing capabilities: 
	\begin{itemize}
	\item {In-chip multiplexing} -- Certain technologies allow to measure several different spectra within a single integrated spectrographs (up to 30). However the performance of additional channels can be reduced, and the cost in size can be high since it generally requires a cross-dispersion (bulk) stage to disentangle spectra. Energy-sensitive detectors could be a solution to preserve compactness.
	\item {3D multiplexing} -- Considering the case of {\it semi}-integrated spectrographs, the compact size can be a disadvantage as very tiny parts have to be aligned at very high precision. Multiplexing capabilities are then limited for such devices. On the other hand, fully integrated technologies allow to easily stack tenths of spectrographs in a very compact and stable fashion.
	\end{itemize}
\item {\bf Operability} -- Instruments can necessitate to be finely tuned or calibrated because of their sensitivity to any position change along its use (e.g.~Fabry-P\'erot are very sensitive to attitude evolution). Fully integrated spectrographs have the advantage to have no moving parts and to provide more stable set-up as it has been demonstrated in optical interferometry for instance. The assembly of the different subsystems is precisely done at manufacturing time and the system size remains small, allowing a secure packaging. 
\item {\bf Spectral Filling Factor $F_F$} -- A spectrum can be considered as a limited bandwidth signal made of $k$ spread lines of equivalent width $W_i$ in a wide spectral range $\Dlambda$. The spectral filling factor $F_F$ characterizes then the quantity of information contained in the spectral bandwidth as $F_F = \frac{\sum_i^n W_i }{\Dlambda}$. It is a property inherent to the source, but the spectrograph should fit to this constraint.
\end{itemize}

\noindent For each group of spectrometer we were able to find in the literature several integrated spectrographs. Tab.~\ref{tab:gratings_comparison} to \ref{tab:FTS-compare} summarize the properties of the identified technologies for each group.

\section{Performance comparison} % (fold)
\label{part:signal_to_noise}

\begin{table*}
\caption[Typical sky background flux in visible and IR. ]{Typical sky background flux in visible and IR \cite{patat_2003a, cuby_2000a}. \label{tab:background}}
\begin{center}
\begin{tabular}{llrrrrrrr}
\hline \hline
Band && V & I & J & H & Ks & L & M\\
\hline
$B_\lambda$ &[mag.arsec$^{-2}$] & 21.6 & 19.6 & 16.5 &14.4 & 13.0 & 3.9 & 1.2 \\
$B_\lambda$ for $S\Omega=\lambda^2$ &[ph.s$^{-1}$] & 0.2 & 4 & $77$ & $730$ & $2\cdot 10^3$ & $9\cdot 10^6$ & $8\cdot 10^7$\\
\hline
\end{tabular}
\end{center}
\end{table*}

We compared the performance of the different technologies in different conditions to assess their domain of best use, or even if some concepts can be discarded from this sole point of view.

\subsection{Signal to noise ratio}

The general expression of the signal-to-noise ratio for spectrographs in a given spectral channel is:
\begin{equation}
\frac{\mbox{S}}{N}= \frac{S_\lambda}{\sqrt{\delta _{det }^{2} + \delta _{phot}^{2} }},
\end{equation}
where:

\begin{itemize}
\item $S_\lambda$ is the measured signal on a given spectral channel, which expresses as 
\begin{equation}
	S_\lambda = \rho  \cdot  T \cdot  F_{\lambda }\cdot  t,
\end{equation}
with $\mathbf{\rho}$ the coupling efficiency ($\leq 1$ for fibered instruments, =1 for bulk ones). $\mathbf{F_{\lambda}}$ denotes the spectral flux density of the target in one spectral channel. $\mathbf{T}$ is the optical throughput of the instrument, including the detector quantum efficiency $\eta$. It also includes fringe contrast in the case of an FTS (due e.g.~to beams imbalance). $\mathbf{t}$ is the integration time.
\item $\mathbf{\delta_{det}}$ is the detector noise linked to detector read-out-noise (RON).
\item $\mathbf{\delta_{phot}}$ is the photon noise, including the source and the background. Tab.~\ref{tab:background} gives typical values of the sky brightness in the optical bands. The sky background increasing in proportion of the observed optical \'etendue, AO are very useful by concentrating the source energy on the smallest possible area of the sky.
%\item $\mathbf{\delta_{resp}}$ is the standard deviation of the pixel response fluctuations, or flat-field noise. It can be considered as an additional noise that applies to the signal plus background as a small fraction of it. The noise is close to percent and is negligible in front of photon or detector noise. It is therefore ignored in the following.
\end{itemize}
This formula has been adapted to take into account the specific source noise of each group of spectrometer:

%%%%%%%%%%%%%%%%%%%%%%%%%%%%%%%%%%%%%%%%%%%%%%%%%%%%%%%%%%%%%%%%%%
%\begin{table}[b]
%\begin{center}
%\caption[SNR of the different concepts.]{SNR of the different concepts. For M-SM and MM instruments, we consider $\rho\sim1$. For the sake of clarity we do not put the photon noise of background in these relations. Nevertheless, we take it into account for computations.
%%\vspace
%\label{tab:SNR-working-conditions}}
%\renewcommand{\arraystretch}{2.5}
%\begin{tabular}{lccc}
%\hline \hline
%& Disperser & FTS & Fabry-P\'erot \\
%\hline
%SM & $\frac{\rho F_\lambda}{\sqrt{\rho F_\lambda + 4 N_{pix} RON^2}}$ & $\frac{\rho F_\lambda}{\sqrt{\rho \Nlambda F_\lambda+ 4\Nlambda RON^2}}$ & - \\
%M-SM & $\frac{F_\lambda}{\sqrt{F_\lambda + 4 N_{pix} N_g RON^2}}$ & $\frac{F_\lambda}{\sqrt{\Nlambda F_\lambda + 4\Nlambda RON^2}}$ & - \\
%MM & $\frac{F_\lambda}{\sqrt{F_\lambda + 4 N_{pix} RON^2}}$ & $\frac{F_\lambda}{\sqrt{\Nlambda F_\lambda + 4\Nlambda RON^2}}$ & $\frac{F_\lambda/\Nlambda}{\sqrt{F_\lambda/\Nlambda + RON^2}}$ \\
%\hline
%\end{tabular}
%\end{center}
%\end{table}
%%%%%%%%%%%%%%%%%%%%%%%%%%%%%%%%%%%%%%%%%%%%%%%%%%%%%%%%%%%%%%%%%%

%%%%%%%%%%%%%%%%%%%%%%%%%%%%%%%%%%%%%%%%%%%%%%%%%%%%%%%%%%%%%%%%%
\begin{figure*}[b]
\centering
\includegraphics[width=0.48\textwidth]{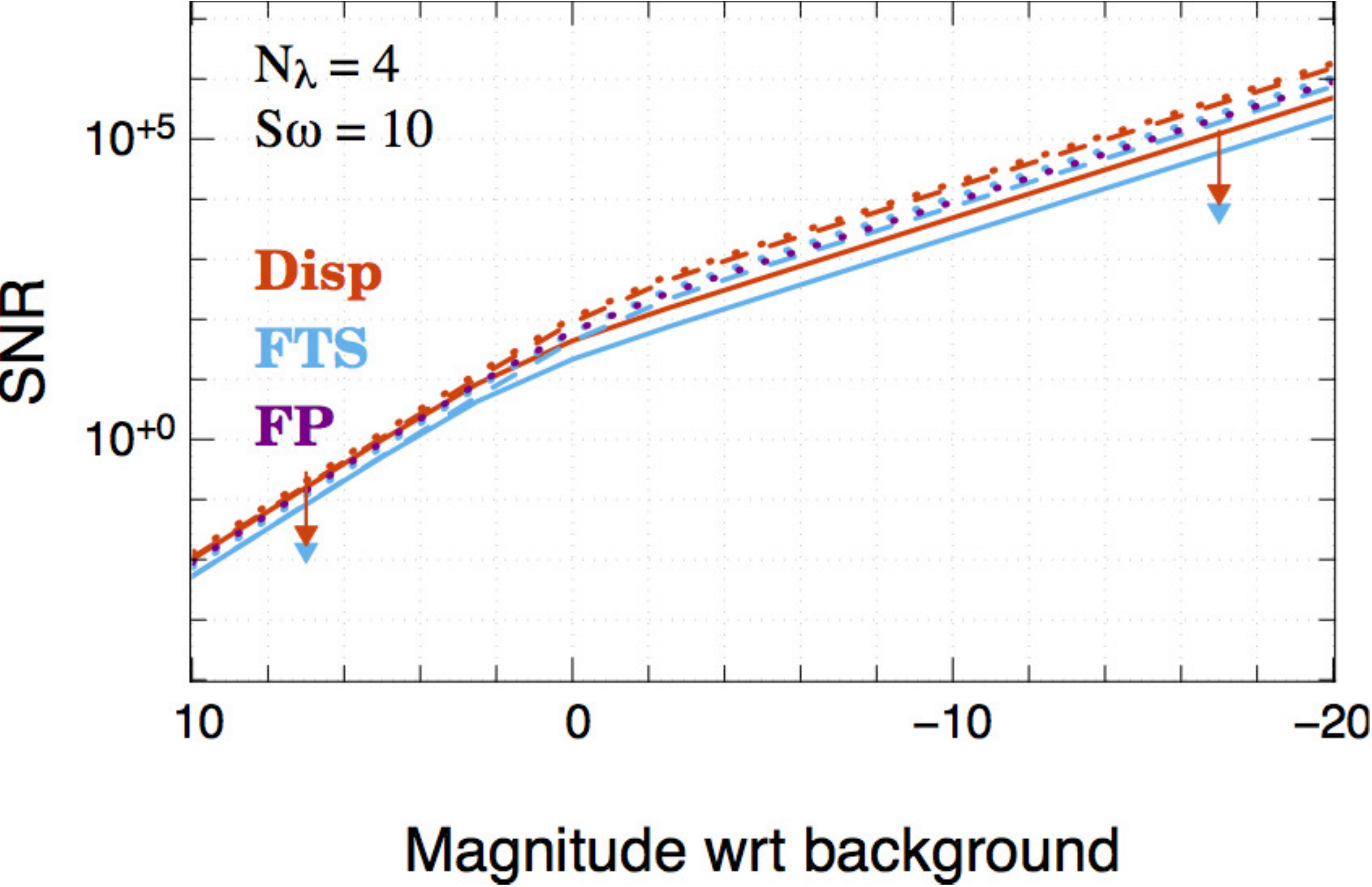}
\hfill
\includegraphics[width=0.48\textwidth]{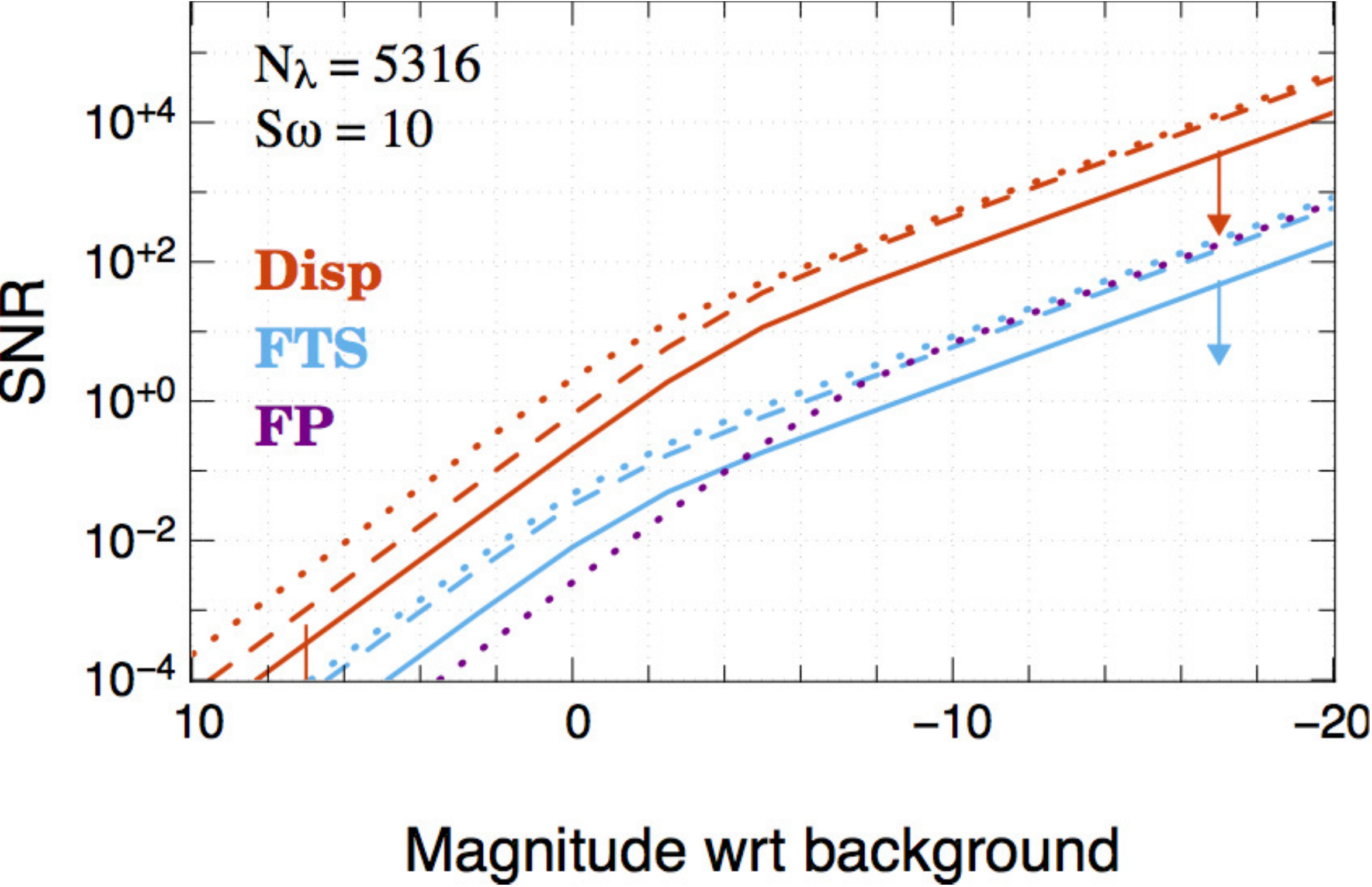}\\
\bigskip
\includegraphics[width=0.48\textwidth]{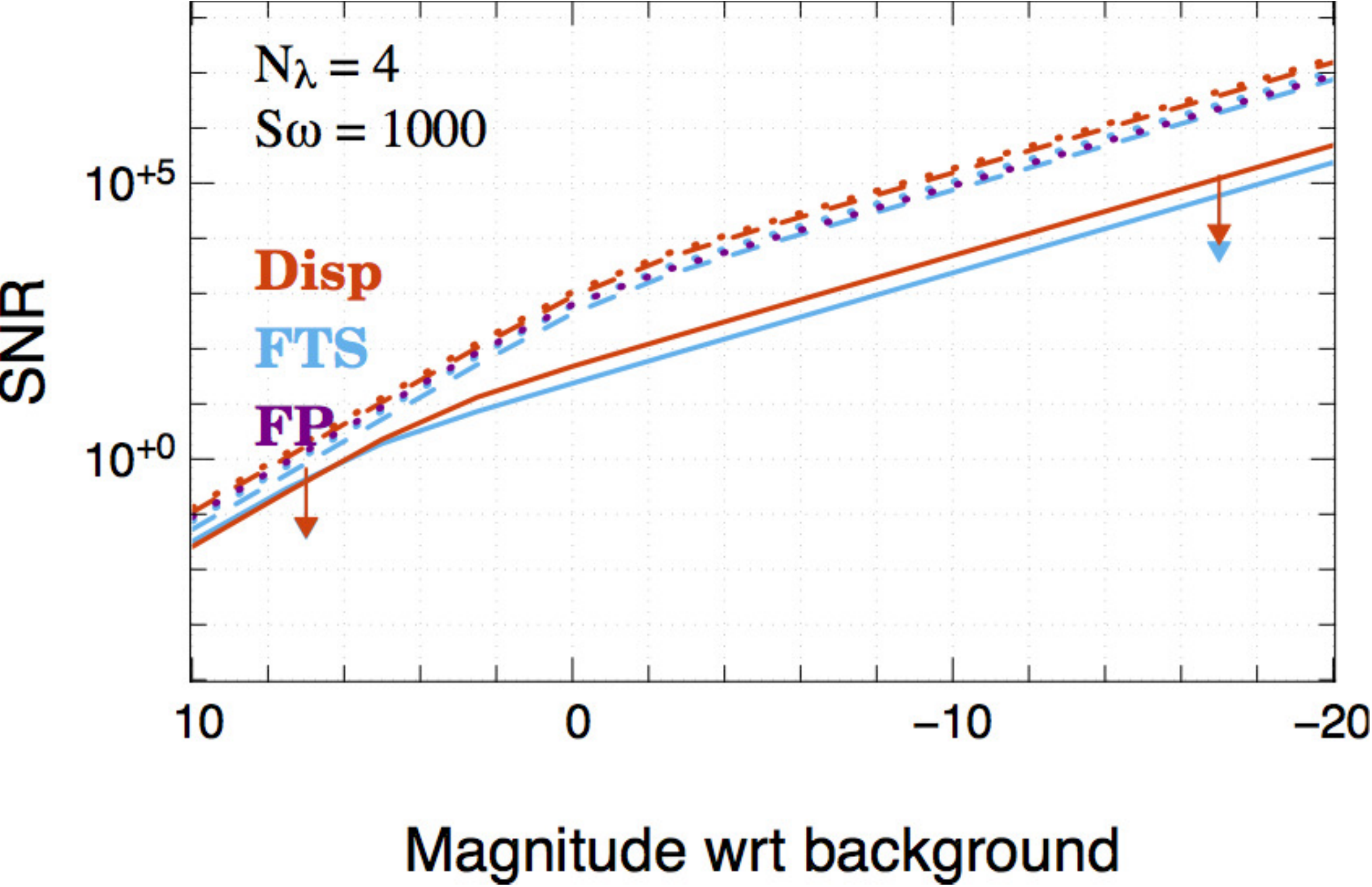}
\hfill
\includegraphics[width=0.48\textwidth]{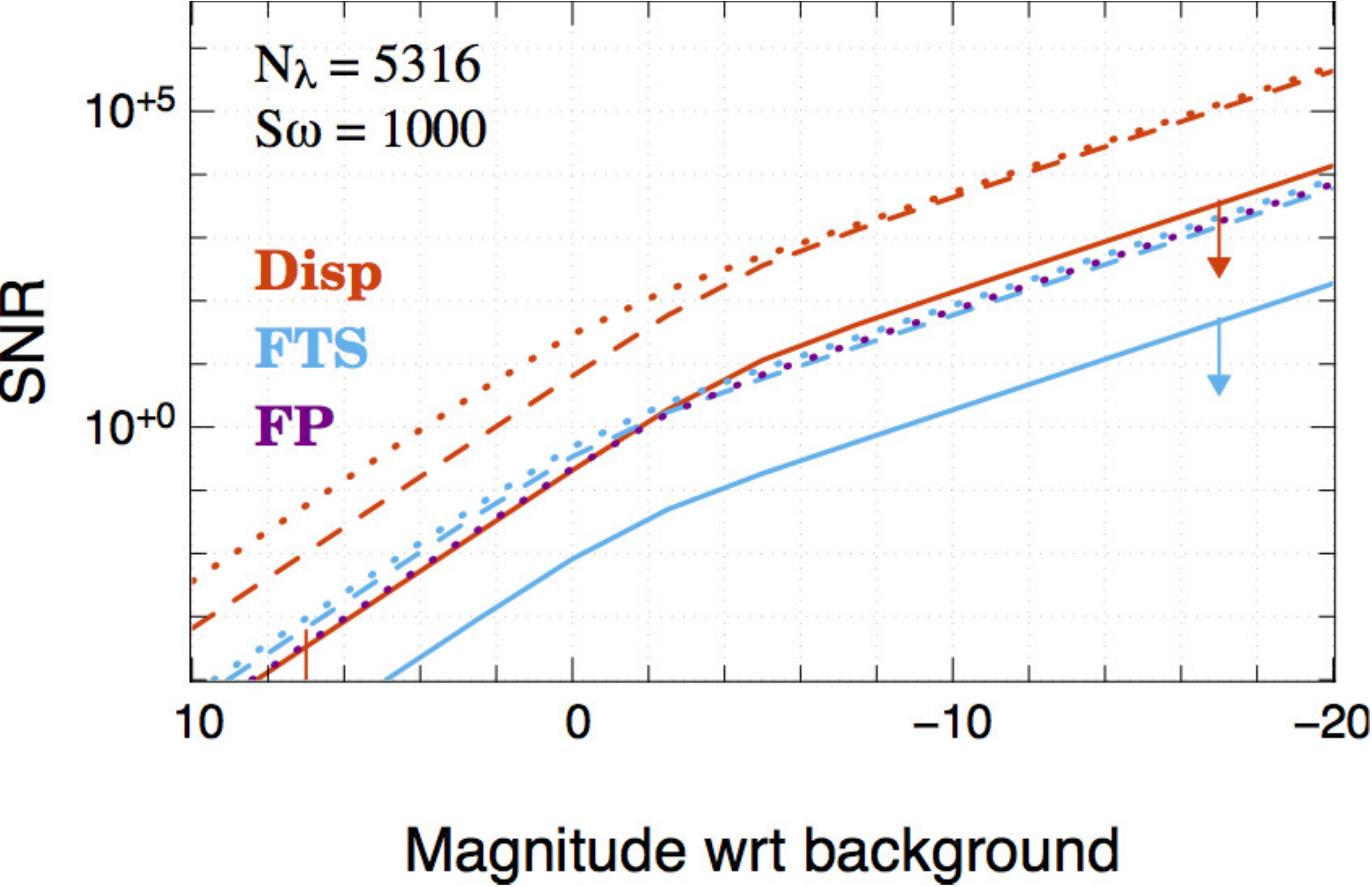}
\caption[SNR vs number of photons for a small source]{SNR vs number of photons for low and high number of spectral channels $N_\lambda$ (left and right). Top and bottom rows correspond to the small and extended source cases ($S\omega = S\Omega/\lambda^2 = M$). We consider SM (solid), M-SM (dash) and MM (dotted) FTS, Dispersers and FP. MM curves are slightly shifted up for the sake of clarity. \label{fig:SNRvsNphot}}
\end{figure*}
%%%%%%%%%%%%%%%%%%%%%%%%%%%%%%%%%%%%%%%%%%%%%%%%%%%%%%%%%%%%%%%%%

\noindent {\bf Disperser --} In a disperser, a pixel is equivalent to a spectral channel. It is then only sensitive to the light from the source $F_\lambda$ and the background $B_\lambda$ at the given wavelength. The flux of a spectral channel is in practice spread over $N_{pix}\sim5$ pixels in the direction perpendicular to the spectra one. The factor 2 in the $B_\lambda$ term accounts for the background subtraction.
\begin{eqnarray*}
&S_\lambda &= \rho  \cdot  T \cdot  F_{\lambda }\cdot  t\\
&\delta_{det}^2 &= N_{pix}  RON^2\\
&\delta_{phot}^2 &= \rho  T t (F_\lambda + 2 B_\lambda)
\end{eqnarray*}

\noindent {\bf Fabry-P\'erot --} In a FP, only one spectral channel is measured at once. To scan the whole spectral band, only $1/N_\lambda$ of the whole integration time can be dedicated to each spectral channel:
\begin{eqnarray*}
&S_\lambda &= \rho  \cdot  T \cdot  F_{\lambda }\cdot  t/N_\lambda\\
&\delta_{det}^2 &= RON^2\\
&\delta_{phot}^2 &= \frac{1}{N_\lambda} \rho  T (F_{\lambda }+2 B_\lambda)  t
\end{eqnarray*}

\noindent {\bf FTS --} Each pixel samples fringes resulting from the interference of the light from the {\it whole} spectral band, and suffers then from an increased photon noise. The Fourier transform then passes the noise accumulated by all the pixels to each spectral channel:
\begin{eqnarray*}
&S_\lambda &= \rho  \cdot  T \cdot  F_{\lambda }\cdot  t\\
&\delta_{det}^2 &= N_\lambda  RON^2\\
&\delta_{phot}^2 &=  N_\lambda\rho  T t (F_{\lambda } + 2B_\lambda)
\end{eqnarray*}

\noindent The coupling efficiency is approximated as the ratio of the instrument to the object \'etendue:
\begin{equation} \label{eq:rho_strehl}
\rho = min(M \lambda^2/S\Omega, 1)
\end{equation}
In the case of a SM instrument observing a point source: $\rho = 0.82 \times \St$ \cite{coudeduforesto_2000}, where $\St$ is the Strehl ratio. %Tab.~\ref{tab:SNR-working-conditions} present simplified overview of the SNR formula for the different cases considered. 
Considering the sky background as the limiting factor (photon noise limited observations), we can compare the performance of the different integrated spectrographs independently of the wavelength.

%%%%%%%%%%%%%%%%%%%%%%%%%%%%%%%%%%%%%%%%%%%%%%%%
\begin{figure}[b]
\centering
\includegraphics[width=0.8\textwidth]{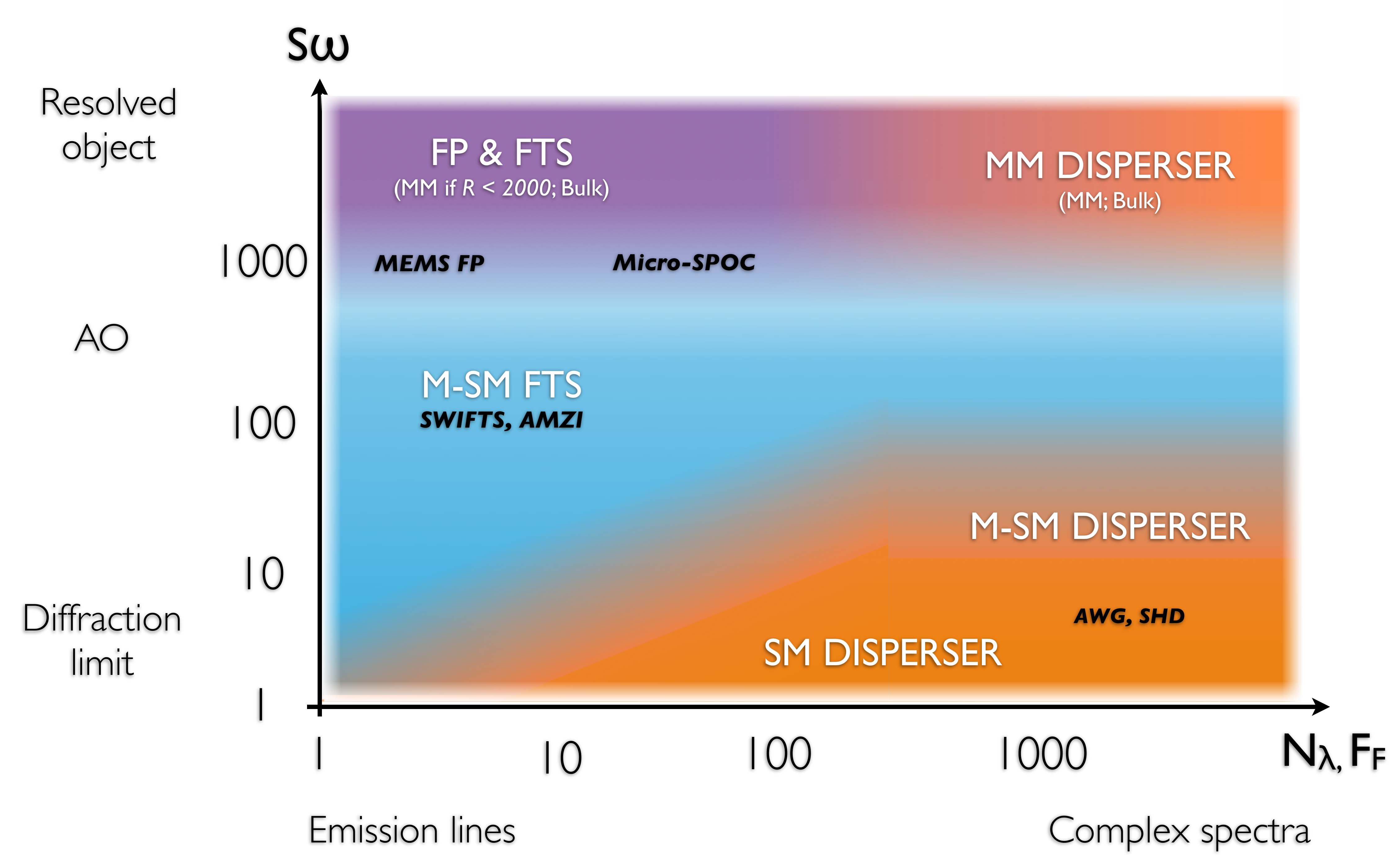}
\caption[Figure of merit of the different kind of instruments in the framework of astrophotonics.]{Figure of merit of the different kind of instruments in the framework of astrophotonics. Following the conclusions of the performance analysis, this figure shows the preferred operating regimes of the different families of spectrographs, depending on the spectrum complexity (linked to the number of spectral channels) and on the source extent. The different areas follow the nomenclature for the performance plots: orange for dispersers, blue (and violet) for FTS, and violet for FP.  We also report in black the identified technologies with the best potential (in our opinion). \label{fig:SpectroSumUp}}
\end{figure}
%%%%%%%%%%%%%%%%%%%%%%%%%%%%%%%%%%%%%%%%%%%%%%%%

\subsection{Analysis of performance}

The performance comparison can be summarized by the four "extreme" cases presented in Fig.~\ref{fig:SNRvsNphot}, consisting of a point or extended source (top and bottom), with low or high number of spectral channels (left and right). We compare only performance of concepts here, and put aside the transmission considerations.
\begin{itemize}
\item It is no surprise that MM instruments (bulk included) show overall better performance thanks to an optimal capacity to collect the telescope light.
\item  In the context of SM spectrographs, M-SM instruments can perform almost as well. They are especially interesting solutions in NIR and even more in MIR wavelengths, where the ELTs wide-field AO systems can provide an excellent correction of the turbulent wavefront (Tab.~\ref{tab:strehl_ELT}). However, they require several individual spectrographs for each spatial elements, and as many times more pixels, also resulting in a higher detector noise which can be problematic for faint object applications. Developments of detectors with rectangular pixels could help on this aspect . The situation with moderately extended sources ($< 100$\: mas) when considering visible wavelengths is more problematic: this requires hundreds to thousands of integrated spectrographs per spatial element. The compactness of such instruments should also be studied.
\item SM devices appear naturally at their disadvantage in the ELT context, excepted if coupled to eXtreme AO systems. More generally, the current technologies have an interest on "few $r_0$" scale telescopes where the \'etendue of the turbulent source is reasonably small. Another field of direct application for SM devices is obviously space missions, providing diffraction limited images.
\end{itemize}

Regarding the families of spectrometers, dispersers are the devices using photons the most efficiently. In the case of small and/or sparse bandwidths however (left plots), FTS and Fabry-P\'erot perform as well: it can be advantageous to consider them in such situations (potentially more sample and less pixel demanding instruments), especially concerning FTS with a dispersed fixed delay configuration for radial velocity measurements. Finally, on the extended objects and high number of spectral channels case (bottom right), it can be noticed that SM dispersers, M-SM FTS and (MM) FP provide all 3 roughly equivalent performance, although not optimal. They would be all the more an appropriate solution than the telescope is small since it suffers less from the atmospheric turbulence and benefits of proportionally smaller beam \'etendue.

 % % % % % % % % % % % % % % % % % % % % % % % % % % % % % % % % % % % % % % % % % % % %
 \section{Conclusion} \label{part:conclusion}

We have presented a non-exhaustive list of integrated spectrograph technologies which are compared on the basis of their estimated performance. The figure~\ref{fig:SpectroSumUp} gives a summary of this analysis, and shows that each technology/type of spectrograph can answer to a specific need. This only gives an overview of the problem of course: making a choice between a technology/concept against another is out of the scope of this study, and require to precisely analyse the constraints of specific science cases.

Efforts are already under way to develop and demonstrate the potential of integrated spectrographs in astrophysics. Examples of such components that have been tested on sky are e.g.~photonic lanterns \cite{horton_2012a} and photonic dicers \cite{harris_2014a}, sometimes in combination with fiber OH-suppression filters \cite{horton_2012a}, as well as integrated spectrographs like the AWG\cite{cvetojevic_2012c}. We believe there is an important case in developing multi-mode solutions: MM spectrographs with a limited but well determined number of modes could be a good trade-off to reach spectral resolutions up to few thousands (relevant for many applications) without limiting the optical \'etendue of the spectrograph. All these photonic devices will also highly benefit from the current detector developments, from rectangular pixels to energy and polarisation sensitive ones, so as to build even more compact and efficient, fully integrated spectrographs.

\acknowledgements{The authors are grateful to the European Union for the funding of the OPTICON Astrophotonics (Work Package 3) and J. Allington-Smith which allowed this study.}

%%%%%%%%%%%%%%%%%%%%%%%%%%%%%%%%%%%%%%%%%%%%%%%%%%%%%%
%%                 REFERENCES                       %%
%%%%%%%%%%%%%%%%%%%%%%%%%%%%%%%%%%%%%%%%%%%%%%%%%%%%%%
\bibliographystyle{spiebib}
\bibliography{SPIE_astrophotonics}   % BiTex files

\end{document}